\documentclass{article}
\usepackage{graphicx} % Required for inserting images
\usepackage{amsmath} % assumes amsmath package installed
\usepackage{amssymb}
\usepackage{authblk}
\title{\bf Towards Streamer-Free Marine Seismic Surveys: The Role of Formation Control Systems}
\author[1]{Saad J. Saleh}
\author[2]{Ed K. Biegert}
\affil[1]{Department of Electrical and Computer Engineering, The University of Texas at Austin, Austin, TX}
\affil[2]{Geophysical Consultant, Houston, TX}
\date{}

\begin{document}

\maketitle

\begin{abstract}

We explore the applicability of recent advances in formation control theory and robust network analysis to the problem of designing streamer-free marine seismic surveys.  To this end, we carry out a theoretical and numerical feasibility study to highlight the power and utility of this methodology, providing an explicit formula for the control law required to continually maneuver the data acquisition formation, and specifically addressing the robustness issue related to potential sensor failures.
\end{abstract}

\section{Introduction}

The development of autonomous vehicles has revolutionized geophysical and geochemical sensing.  Many of the new emerging applications require deployment of a large fleet of unmanned vehicles moving collectively in a formation constrained to maintain a fixed geometric shape while maneuvering.  Such a restriction can be particularly challenging for underwater platforms where GPS signals are unavailable and alternative localization strategies can be rather costly or inaccurate.  

\bigskip
\noindent
The advent of Formation Control Theory (e.g., see Oh et al., 2015) in recent years has provided a theoretical foundation for understanding and controlling formations through simple localized inter-vehicle exchange of information.  Our main motivation in this paper is to conduct a theoretical and numerical feasibility study to explore applicability of formation control theory to cable-free marine seismic survey design, with particular emphasis on robustness considerations.  To this end, we consider a typical marine seismic survey layout, focus on the possibility of replacing the cable-linked receivers by sensor-carrying autonomous vehicles, and address the questions of inter-vehicle sensing and control algorithms required to robustly maintain the desired acquisition geometry while maneuvering.  Our approach is distinguished from earlier seismic-related efforts in this line of research (e.g., see Muyzert, 2018; Mancini et al., 2019; and references therein) in that we provide an explicit formula for the control law required to continually maneuver the formation, and we specifically address the robustness issue related to potential sensor failures.  While we focus primarily on analytical and numerical aspects of the feasibility study, we note that practical implementation of this approach carries a number of additional technical challenges that are the subject of ongoing research.  These challenges are not covered in this short paper.

%%%%%%%%%%%%%%%%%%%%%%%%%%%%%%%%%%%%%%%%%%%%%%%%%%%%%%%%%%%%%%%%%%%%%%%%%%%%%%%%%%%%%%

\section{Problem Formulation}
\label{problem_formulation}

We use the words vehicle, sensor, node, and agent interchangeably to denote the autonomous vehicle and its payload collectively.  We assume the vehicles are fully actuated and each can be represented as a point mass.  Furthermore, we suppose the vehicles have no access to their absolute positions and velocities in a global coordinate system, but can measure their \textit{relative} positions and \textit{relative} velocities with respect to their neighbors (a small subset of the fleet) via short-range simple interactions.

\bigskip
\noindent
Let $n$ be the number of nodes in the fleet, and let $p_i(t) \in \Bbb{R}^3$ and $v_i(t) \in \Bbb{R}^3$ be the position and velocity, respectively, of the $i$th node at time $t \geq 0$.  Assuming unit mass for each vehicle, the equations of motion for the system are $\dot{p_i}(t) = v_i(t)$ and $\dot{v_i}(t) = u_i(t)$ for $i = 1, \ldots, n$, where $u_i(t) \in \Bbb{R}^3$ is the net force acting on the $i$th node at time $t$.  Moreover, we assume the desired formation shape to be tracked by the fleet is pre-specified as $p_i^*(t) \in \Bbb{R}^3$ and $v_i^*(t) \in \Bbb{R}^3$ for each node $i$ and for each $t \geq 0$.

\bigskip
\noindent
Since each vehicle can communicate with only a small subset of the fleet, it is convenient to view the collection of vehicles as a graph $\mathcal{G} = ( \mathcal{V},\mathcal{E} )$ where $\mathcal{V} = \{ 1,\ldots,n \}$ is the set of node labels and $\mathcal{E} \subseteq \mathcal{V} \times \mathcal{V}$ is the set of graph edges containing all vehicle pairs $(i , j)$ that can exchange relative positions and velocities.  We assume the graph is undirected so that $(i,j) \in \mathcal{E} \iff (j,i) \in \mathcal{E}$.  As such, the set of neighbors for each node $i$ is defined as $\mathcal{N}_i \doteq \{ j \in \mathcal{V} : (i,j) \in \mathcal{E} \}$.  Once $\mathcal{N}_i$ is specified for each node $i$, we have a fixed network topology for vehicle communication.

\bigskip
\noindent
\textit{Problem 1 (Formation Control)}:  Design a controller for the force $u_i(t) \in \Bbb{R}^3$ to be applied to each vehicle $i \in \{ 1,\dots,n \}$ as a function of the \textit{relative} errors in position and velocity between node $i$ and its neighbors in $\mathcal{N}_i$ such that
\begin{equation}
   p_i(t) \to p_i^*(t), \quad v_i(t) \to v_i^*(t), \quad \text{as} \quad t \to \infty \quad \text{for} \quad i=1,\ldots,n \; .\label{objective}
\end{equation}

\bigskip
\noindent
To introduce the robustness challenge, we note that \textit{Problem 1} can be solved by using a number of possible network topologies and address the problem of choosing a network that is robust in the face of node failures.  To this end, we consider a parameterized set of graphs motivated by a typical marine seismic survey geometry.  Since inline distances are typically much shorter than crossline distances, we adopt a class of topologies where each vehicle can interact with its two nearest neighbors in the inline direction, but only a fraction $r$ of the vehicles can exchange information with their nearest neighbors in the crossline direction.  Now, taking $f$ to be the fraction of sensors that may fail randomly, we ask whether the surviving vehicles can maintain network integrity.  To be specific, we take $P_{\infty}$ as a proxy for network integrity, where $P_{\infty}$ is the probability that a randomly chosen surviving node belongs to the largest cluster in the surviving network (e.g., see Barabasi, 2016) and formulate the following problem.

\bigskip
\noindent
\textit{Problem 2 (Robust Network Topology)}:  For any $f$ chosen to approximate the expected fraction of node failures, compute the required fraction of nodes $r$ that need to exchange information with their nearest neighbors in the \textit{crossline} direction to guarantee that $P_{\infty} \geq 0.9$.

%%%%%%%%%%%%%%%%%%%%%%%%%%%%%%%%%%%%%%%%%%%%%%%%%%%%%%%%%%%%%%%%%%%%%%%%%%%%%%%%%%%%%%

\section{Results: Formation Control}
\label{displacement_control}

\begin{figure}
\begin{minipage}[c]{0.4\linewidth}
   \centering
   \includegraphics[width=5.25cm]{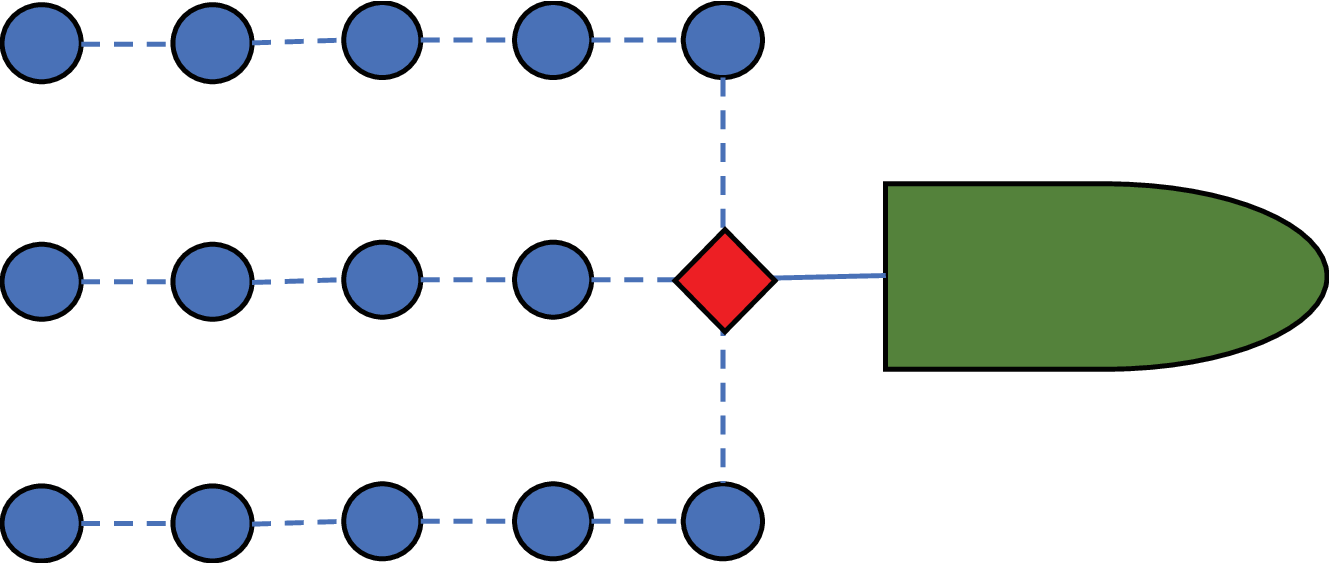}
   \caption{Network for Miniature Survey (with "leader" vehicle as red diamond).}
   \label{fig:survey_with_leader}
\end{minipage} \hfill
\begin{minipage}[c]{0.4\linewidth}   
   \centering
   \includegraphics[width=5.25cm]{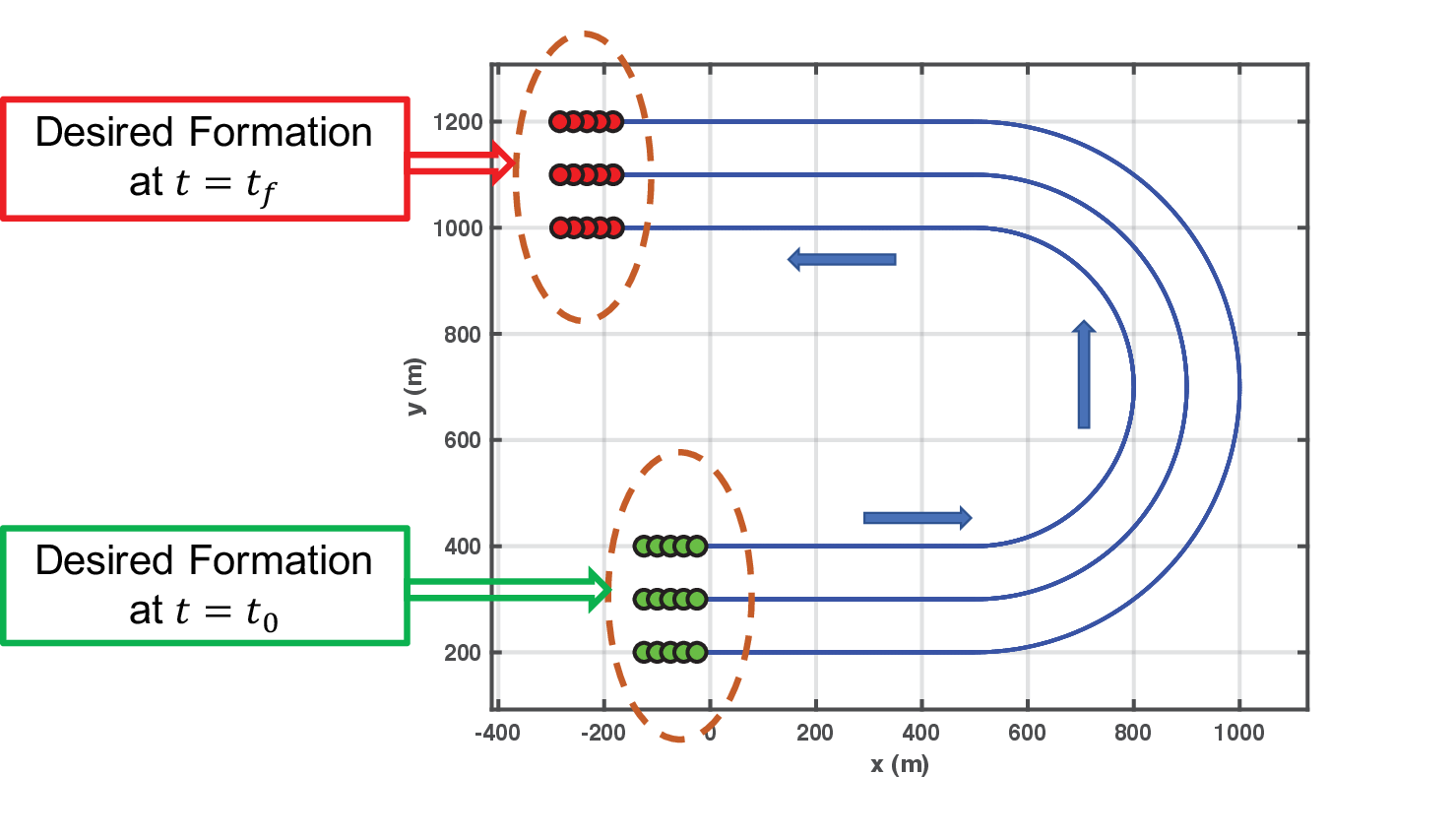}
   \caption{Desired Trajectories.}
   \label{fig:desired_trajectories}
\end{minipage}
\end{figure}

\bigskip
\noindent
In general, the design objective specified by (\ref{objective}) cannot be met if none of the vehicles can sense its absolute position and velocity.  Hence, we adopt a leader-following strategy and designate one vehicle as a "leader", which in principle can be towed by a boat.  All others are autonomous and rely on inter-vehicle communication.  Now, for a network where each node can communicate only with its two nearest inline neighbors (except for the first ``column'' containing the leader, where crossline communication is allowed --- see Figure \ref{fig:survey_with_leader}), we define the relative position and velocity signals for the $i$th node and its $j$th neighbor as $\Delta_{pij} \doteq p_j - p_i$, $\Delta_{vij} \doteq v_j - v_i$, respectively, and (following Hong et al., 2006 and Ren and Atkins, 2007) use the force control law
\begin{equation}
   u_i = g_p \sum_{j \in \mathcal{N}_{i}} ( \Delta_{pij} - \Delta_{p^*ij} )
+ g_v \sum_{j \in \mathcal{N}_{i}} ( \Delta_{vij} - \Delta_{v^*ij} ) + \delta_{il} g_p (p_l^* - p_l) + \delta_{il} g_v (v_l^* - v_l) , \label{leader_displacement_controller}
\end{equation}
for $i = 1,\ldots,n$, where $p_l$ and $v_l$ denote the leader's position and velocity, respectively, $\delta_{il} = \delta (i-l)$ is the Kronecker delta function, and $g_p, g_v > 0$ are user-specified gain functions.  Now, with $p \doteq [ p_1^T \ldots p_n^T ]^T$, $v \doteq [ v_1^T \ldots v_n^T ]^T$ (and similarly for $p^*$ and $v^*$), and with error vectors defined as $e_p \doteq p^* - p$, $e_v \doteq v^* - v$, the system's error dynamics after applying the control laws $u_i$ are given by 
\begin{eqnarray}
   \dot{e} & = & \Gamma \; e , \label{error_dynamics_with_leader} \\
   \begin{bmatrix} \dot{e_p} \\ \dot{e_v} \end{bmatrix} & = & \begin{bmatrix} 0_{3n} & I_{3n} \\ -g_p[(L+H) \otimes I_3] & -g_v[(L+H) \otimes I_3] \end{bmatrix}  \begin{bmatrix} e_p \\ e_v \end{bmatrix} ,  \nonumber
\end{eqnarray}
where $H = \text{diag}(\delta_{1l}, \ldots, \delta_{nl})$, $L$ is the Laplacian matrix of the underlying graph, $0$ and $I$ are square zero and identity matrices of dimensions indicated by their subscripts, and $\otimes$ is the matrix Kronecker product.  To achieve the desired objective in (\ref{objective}), $\Gamma$ is required to be a Hurwitz matrix; i.e., the real parts of all its eigenvalues should be negative, thus ensuring the error is driven towards zero, $e(t) \to 0$ as $t \to \infty$.  This is indeed the case for proper choices of the design parameters $g_p$ and $g_v$, as illustrated in the following example.

\bigskip
\noindent
\textit{Example 1}:  Consider the miniature seismic survey shown in Figure \ref{fig:survey_with_leader}, consisting of 15 nodes.  Assume the inline distance between adjacent nodes is 25~m and the crossline distance is 100~m.  Our objective is to move the fleet from a desired initial formation at $t_0 = 0$ to a desired final formation at $t_f = 750$~seconds along pre-specified trajectories as shown in Figure~\ref{fig:desired_trajectories}.  To achieve this objective, we invoke control law (\ref{leader_displacement_controller}) to simulate the formation dynamics.  Choosing $g_p = g_v = 0.5$ in (\ref{leader_displacement_controller}), the eigenvalues of the error dynamics matrix $\Gamma$ can readily be computed and indeed all have negative real parts, with 
\begin{equation*}
   -1.4606 \leq \text{real} \, \{ \lambda_k(\Gamma) \} \leq -0.0137 \quad \text{for} \;\; k=1,\ldots,60 .
\end{equation*}
This ensures that the tracking objective (\ref{objective}) is achieved.  

%%%%%%%%%%%%%%%%%%%%%%%%%%%%%%%%%%%%%%%%%%%%%%%%%%%%%%%%%%%%%%%%%%%%%%%%%%%%%%%%%%%%%%

\section{Results: Robust Topologies}
\label{robustness}

The leader-following control law discussed in the previous section maintains the desired formation as long as the network remains connected.  We now turn to the question of robustness and explore the ability of the formation to maintain some level of integrity in the face of random node failures.  Clearly, the larger the fraction $f$ of nodes that randomly fail, the more fragmented the surviving network.  Our goal is to analyze integrity of the surviving network as a function of $f$.  This problem has been extensively studied in the last 20 years (e.g., see Callaway et al., 2000, Cohen et al., 2000, and Barabasi, 2016) but to the best of the authors' knowledge, never applied to the formation control problem considered here.

\begin{figure}
\begin{minipage}[c]{0.4\linewidth}
   \centering
   \includegraphics[width=5.25cm]{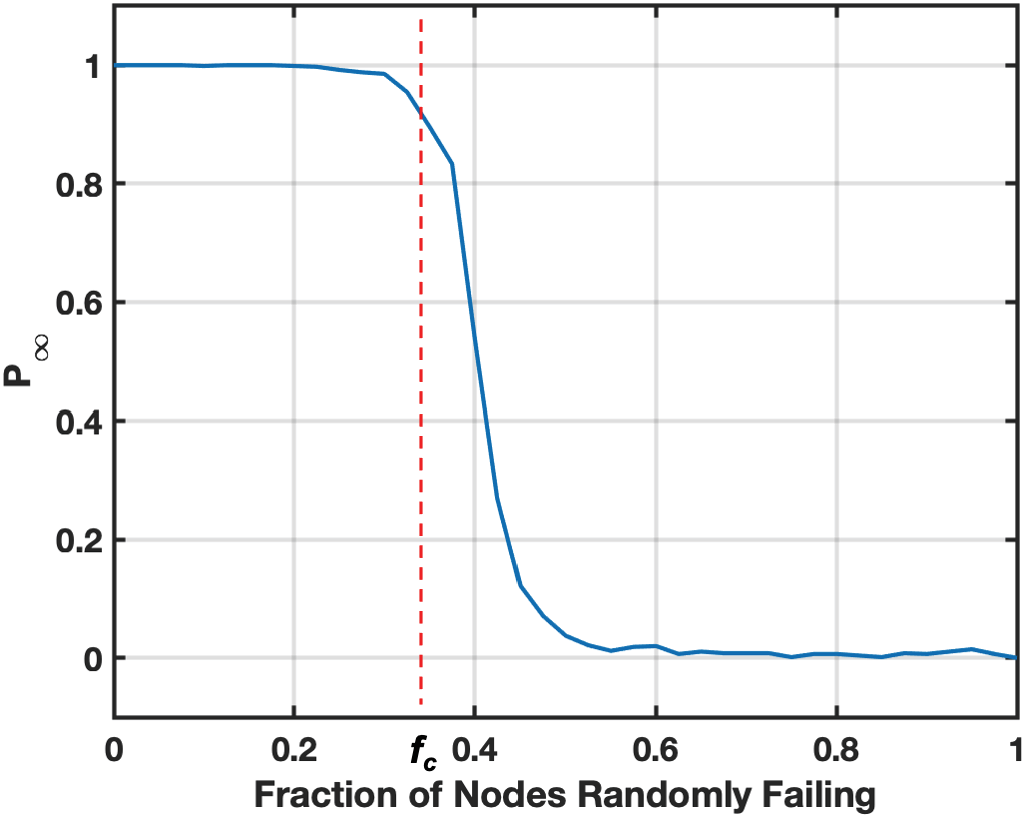}
   \caption{$P_{\infty}$ as a function of $f$ for a 2D square lattice.}
   \label{fig:P_inf}
\end{minipage} \hfill
\begin{minipage}[c]{0.4\linewidth}   
   \centering
   \includegraphics[width=5.25cm]{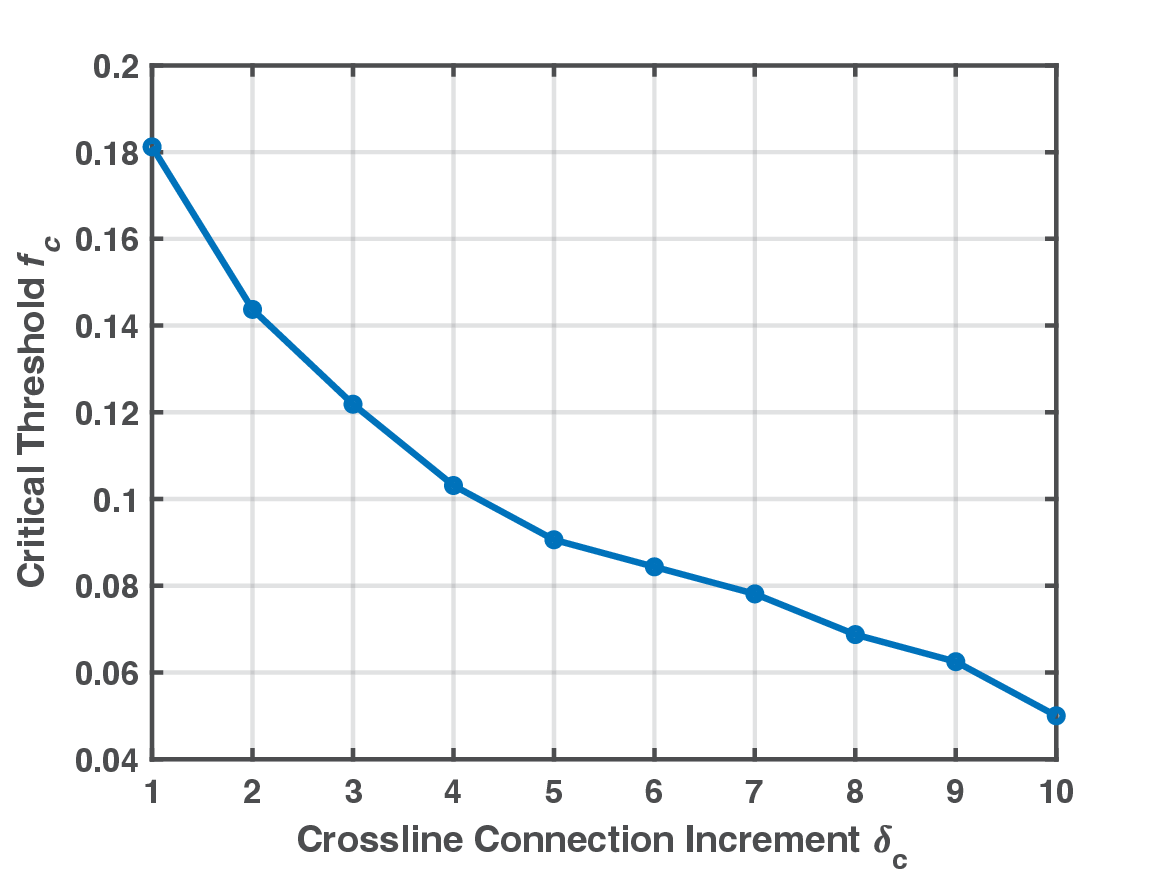}
   \caption{Critical Threshold $f_c$ as a Function of Crossline Connectivity Increment $\delta_c$.}
   \label{fc_versus_delc}
\end{minipage}
\end{figure}

\bigskip
\noindent
One commonly used metric for characterizing and quantifying the notion of integrity for the surviving network is the order parameter $P_{\infty}$ defined in the \textit{Problem Formulation} section.  Clearly, $P_{\infty}(f)$ is expected to be a decreasing function, but interestingly it often decreases nonlinearly with a characteristic shape that resembles a phase transition.  To illustrate, Figure~\ref{fig:P_inf} shows $P_{\infty}(f)$ when the underlying network is a generic 2-dimensional square lattice.  The phase transition occurs at a critical threshold $f_c$ of node failures.  As such, one expects the surviving network to remain largely effective as long as $f < f_c$, and to largely disintegrate when $f > f_c$.  Hence, a sensible design strategy is to choose a network topology that maximizes $f_c$.  However for non-random (correlated) networks, such as the ones considered in this paper, no simple formula for $f_c$ exists, and $f_c$ must be computed numerically.  For our application, we take $f_c$ such that $P_{\infty}(f_c) = 0.9$.

\bigskip
\noindent
Returning to the seismic survey-inspired topologies discussed in the previous sections, we allow each node to exchange information with its two nearest neighbors in the inline direction but limit crossline communication to the smallest fraction $r \in [0,1]$ of nodes that guarantees a certain level of network integrity.  This is illustrated in the following example.

\bigskip
\noindent
\textit{Example 2}:  Consider a sensor layout for a seismic survey consisting of 250 nodes arranged in 5 rows, with 50 nodes in each row.  We take the inline distance to be 25 m and the crossline distance 100 m.  Our objective is to move this rectangular formation along straight 2,250-meters-long trajectories.  This setup is similar to Example~1, albeit much larger, and hence we expect application of control law (\ref{leader_displacement_controller}) to achieve the desired objective as long as the network remains connected.

\bigskip
\noindent
To consider the robustness question, we define $\delta_c \doteq 1/r$ as the node increment for crossline communication and examine $f_c$ as a function of $\delta_c$.  For example, $r=0.1$ implies that every 10\textsuperscript{th} node is allowed to communicate with its nearest crossline neighbors ($\delta_c = 10$) and $r = 1/3$ implies every 3\textsuperscript{rd} node is allowed to communicate with its nearest crossline neighbors ($\delta_c = 3$).  Figure~\ref{fc_versus_delc} shows the dependence of $f_c$ on $\delta_c$, generated via a Monte Carlo scheme that simulated 1500 trials of node failures for each $\delta_c \in \{1,2,\dots,10\}$.  For example, the figure shows that if the expected maximum number of node failures is 5\% ($f_c = 0.05$) then allowing every 10\textsuperscript{th} node to communicate in the crossline direction is sufficient to maintain fleet integrity ($\delta_c = 10$), but if the expected maximum number of node failures is 12\% then crossline communication needs to be considerably more frequent, roughly at the rate of every 3\textsuperscript{rd} node.

%%%%%%%%%%%%%%%%%%%%%%%%%%%%%%%%%%%%%%%%%%%%%%%%%%%%%%%%%%%%%%%%%%%%%%%%%%%%%%%%%%%%%%

\section{Conclusions}
\label{conclusion}

Autonomous vehicles are now widely applied for geophysical and geochemical sensing, but their use as cooperative formations of sensors remains limited.  In this paper, we applied recent results from formation control theory to demonstrate analytically and numerically the feasibility of using a fleet of autonomous vehicles as a formation of receivers in streamer-free marine seismic surveys.  We considered various network topologies and applied robust network theory to guide selection of resilient topologies capable of achieving the data acquisition task in the face of random node failures.  Potential future research directions to expand this work include numerical analysis of under-actuated nonlinear vehicle dynamics, incorporating communication time delays and collision avoidance capabilities, as well as a field trial using appropriate vehicle platforms.

%%%%%%%%%%%%%%%%%%%%%%%%%%%%%%%%%%%%%%%%%%%%%%%%%%%%%%%%%%%%%%%%%%%%%%%%%%%%%%%%%%%%%%

%\section{Acknowledgements (Optional)}

%This is the first sentence of the acknowledgements.

%%%%%%%%%%%%%%%%%%%%%%%%%%%%%%%%%%%%%%%%%%%%%%%%%%%%%%%%%%%%%%%%%%%%%%%%%%%%%%%%%%%%%%

\section{References}

Barabasi, A-L [2016] \textit{Network Science}, Cambridge University Press: Cambridge.

\bigskip
\noindent
Callaway, D.S., Newman, M.E.J, Strogatz, S.H. and Watts, D.J. [2000] Network robustness and fragility: percolation on random graphs.  \textit{Physical Review Letters}, \textbf{85}, (25), 5468-5471.

\bigskip
\noindent
Cohen, R., Erez, K., ben-Avraham, D. and Havlin, S. [2000] Resilience of the internet to random breakdowns. \textit{Physical Review Letters}, \textbf{85}, (21), 4626-4628.

\bigskip
\noindent
Hong, Y., Hu, J. and Gao, L. [2006] Tracking control for multi-agent consensus with an active leader and variable topology. \textit{Automatica}, \textbf{42}, 1177-1182.

\bigskip
\noindent
Mancini, F., Debens, H. and Hollings, B. [2019] Low-powered autonomous underwater vehicles for large-scale ocean-bottom acquisition. \textit{First Break}, \textbf{37}, 49-55.

\bigskip
\noindent
Muyzert, E. [2018]. Design, modeling and imaging of marine seismic swarm surveys. \textit{Geophysical Prospecting}, \textbf{66}, 1535-1547.

\bigskip
\noindent
Oh, K-K, Park, M-C and Ahn, H-S [2015] A survey of multi-agent formation control. \textit{Automatica}, \textbf{53}, 424-440.

\bigskip
\noindent
Ren, W. and Atkins, E. [2007] Distributed multi-vehicle coordinated control via local information exchange. \textit{International Journal of Robust and Nonlinear Control}, \textbf{17}, 1002-1033.

\end{document}